\documentclass[conference]{IEEEtran}
\pdfoutput=1

\usepackage{cite}
\usepackage{amsmath,amssymb,amsfonts}
\usepackage{algorithmic}
\usepackage{graphicx}
\usepackage{orcidlink}
\usepackage{textcomp}
\usepackage{xcolor}

\usepackage{glossaries}
\glsdisablehyper

\newacronym{ai}{AI}{artificial intelligence}
\newacronym{cpcl}{CPCL}{cooperative passive coherent location}
\newacronym{doa}{DoA}{direction of arrival}
\newacronym{dod}{DoD}{direction of departure}
\newacronym{dl}{DL}{down link}
\newacronym{gbscm}{GBSCM}{geometrically based statistic channel model}
\newacronym{gnb}{gNB}{gNodeB}
\newacronym{isac}{ISAC}{integrated sensing and communication}
\newacronym{los}{LoS}{line-of-sight}
\newacronym{mec}{MEC}{mobile edge cloud}
\newacronym{mimo}{MIMO}{multiple-input multiple-output}
\newacronym{ml}{ML}{machine learning}
\newacronym{ofdma}{OFDMA}{orthogonal frequency-division multiple access}
\newacronym{ofdm}{OFDM}{orthogonal frequency-division multiplexing}
\newacronym{otfs}{OTFS}{orthogonal time frequency space}
\newacronym{rcs}{RCS}{radar cross section}
\newacronym{rru}{RRU}{remote radio unit}
\newacronym{rx}{Rx}{receiver}
\newacronym{snr}{SNR}{signal-to-noise ratio}
\newacronym{tdma}{TDMA}{time-division multiple access}
\newacronym{tof}{ToF}{time of flight}
\newacronym{tx}{Tx}{transmitter}
\newacronym{ue}{UE}{user equipment}
\newacronym{ul}{UL}{up link}
\newacronym{vna}{VNA}{vector network analyzer}

\usepackage{siunitx}
\DeclareSIUnit{\mms}{\milli\squaremetre}
\DeclareSIUnit{\inch}{in}
\DeclareSIUnit{\inchs}{in\squared}
\DeclareSIUnit{\mil}{mil}
\DeclareSIUnit{\Msps}{Msps}
\DeclareSIUnit{\Mbps}{Mbps}
\DeclareSIUnit{\LSB}{LSB}
\DeclareSIUnit{\pFS}{\percent FS}
\DeclareSIUnit{\dBc}{\deci\bel c}
\DeclareSIUnit{\dBm}{\deci\bel m}
\DeclareSIUnit{\dBFS}{\deci\bel FS}
\DeclareSIUnit{\dBi}{\deci\bel i}
\DeclareSIUnit{\hex}{0x}
\DeclareSIUnit{\vp}{\volt_{\text{p}}}
\DeclareSIUnit{\vpp}{\volt_{\text{pp}}}
\DeclareSIPrefix{\micro}{\text{\textmu}}{-6}

\title{Characterization of Multi-Link Propagation and Bistatic Target Reflectivity for Distributed Multi-Sensor ISAC}

\author{%
    \IEEEauthorblockN{%
        Reiner S. Thomä\,\orcidlink{0000-0002-9254-814X},
        Carsten Andrich\,\orcidlink{0000-0002-4795-3517},
        Julia Beuster\,\orcidlink{0000-0003-1887-4278},
        Heraldo Cesar Alves Costa\,\orcidlink{0009-0002-6186-5780},\\
        Sebastian Giehl\,\orcidlink{0009-0008-1672-1351},
        Saw James Myint\,\orcidlink{0009-0007-3788-7126},
        Christian Schneider\,\orcidlink{0000-0003-1833-4562},
        Gerd Sommerkorn\,\orcidlink{0009-0003-1111-322X}
    }
    \IEEEauthorblockA{%
        \textit{Electronic Measurement and Signal Processing Research Group} \\
        \textit{Technische Universität Ilmenau, Germany}\\
        reiner.thomae@tu-ilmenau.de
    }
}

\begin{document}

\maketitle

\begin{abstract}
\Gls{isac} qualifies mobile radio systems for detecting and localizing of passive objects by means of radar sensing.
Advanced \gls{isac} networks rely on meshed mobile radio nodes (infrastructure access and/or user equipment, resp.) establishing a distributed, multi-sensor MIMO radar system in which each target reveals itself by its bistatic backscattering. Therefore, characterization of the bistatic reflectivity of targets along their trajectories of movement is of highest importance for \gls{isac} performance prediction.
We summarize several challenges in multi-link modeling and measurement of extended, potentially time-variant radar targets.
We emphasize the specific challenges arising for distributed \gls{isac} networks and compare to the state of the art in propagation modeling for mobile  communication. 
\end{abstract}


\section{Introduction}

\Gls{isac} is considered to be one of the key features of future 6G mobile radio.
Despite of different interpretations, we understand \gls{isac} as a means of radar detection and location of passive objects (``targets") that are not equipped with a radio tag.
These targets reveal their existence and position by radio wave reflection only when properly illuminated.
In contrast to well-known radar systems, \gls{isac} exploits the inherent resources of the mobile radio system on both the radio access and network level.
In its most resource efficient operational mode, \gls{isac} reuses the signals originally transmitted for communication purposes at the same time also for target illumination.
This scheme resembles and extends the well-known passive radar principle.
We introduced the term ``\gls{cpcl}" \cite{CPCLTho2019}, \cite{JCASOverviewTho2021} for it.
In case of this communication centric version of \gls{isac}, the radio access modes defined for communication are also used to radar sensing.
This includes the waveform (usually OFDM and derivatives), its numerology, multiuser access (OFDMA, TDMA), pilot schemes, channel state estimation and synchronization, channel state signaling for predistortion and link adaptation, and eventually also for resource allocation.
With the ubiquitous availability of the mobile radio access, we immediately have a distributed network of radar sensors at hand.
The same network is also used for data transport and data fusion.
With the computing facilities of the \gls{mec} we have all resources at our disposal, which we may need to apply \gls{ml} and \gls{ai} for adaptive resource allocation, target parameter estimation, and scene recognition.
This way, \gls{isac} will become a ubiquitous and cognitive radar sensing network. 
As we know very well from mobile radio performance prediction, the knowledge about the multipath radio propagation is very important.
Channel measurement and modeling always stands at the very beginning of the definition and standardization of new radio access schemes.
In this paper, we ask the question: ``What are the differences and challenges of propagation research for \gls{isac} as compared to plain mobile radio communication?"
We will find out among other things that the knowledge about single, i.e., solitaire objects that are identified as radar targets, is most important.
This includes bistatic target reflectivity, how it evolves if the target is moving, and how it can be characterized if it is inherently time-variant.
Besides of conceptual issues, we for the first time introduce a new measurement range for the bistatic reflectivity of extended objects up to the size of a passenger car.
This unique measurement range, which we call BiRa (Bistatic Radar), is capable of real-time wideband measurements of time-variant targets.
Hence, we can analyze the bistatic micro-Doppler response of extended targets~\cite{MicZha2017}.

\section{Multi-link ISAC system issues}
\begin{figure}[t]
    \centering
    \includegraphics[width=\columnwidth]{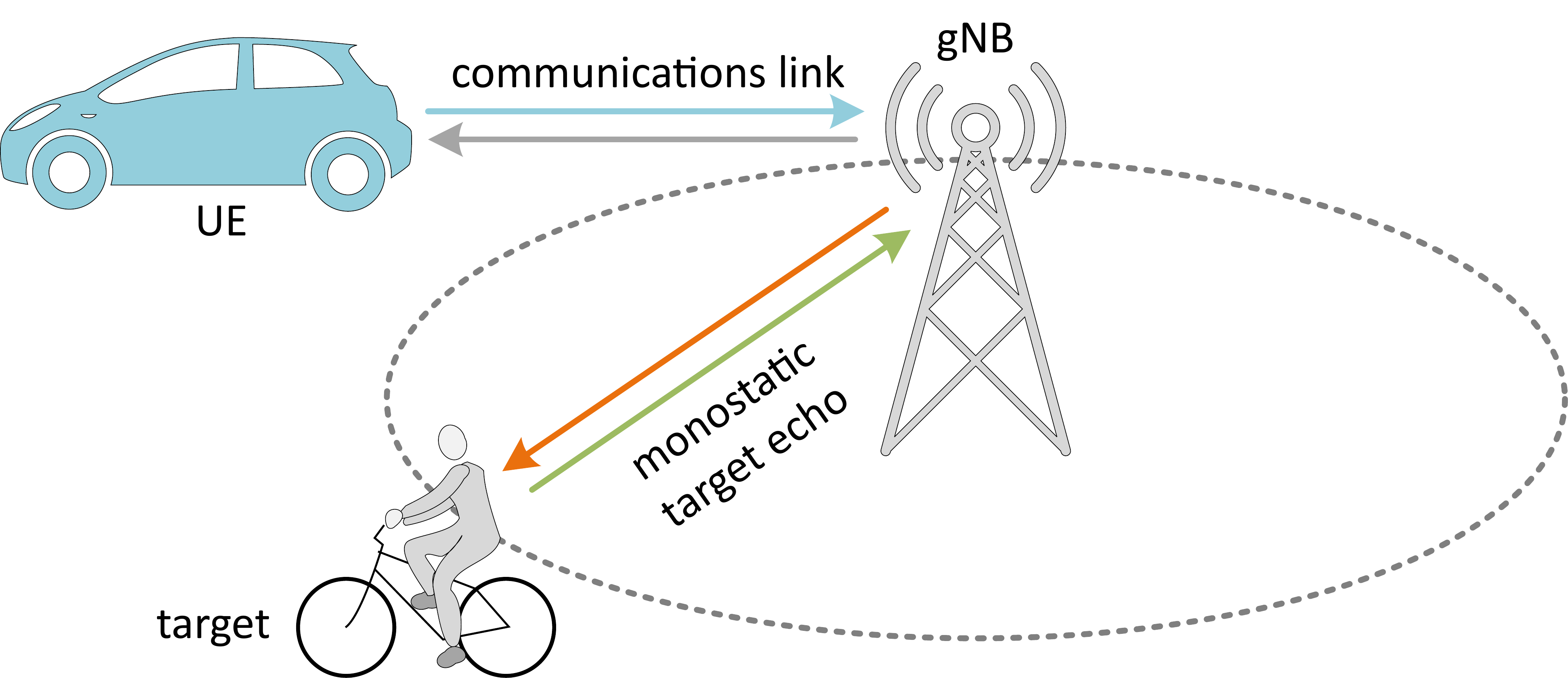}
    \caption{Infrastructure based sensing using a single base station that is equipped with an antenna array.}
    \label{fig:monostaticing}
\end{figure}

\begin{figure}[t]
    \centering
    \includegraphics[width=\columnwidth]{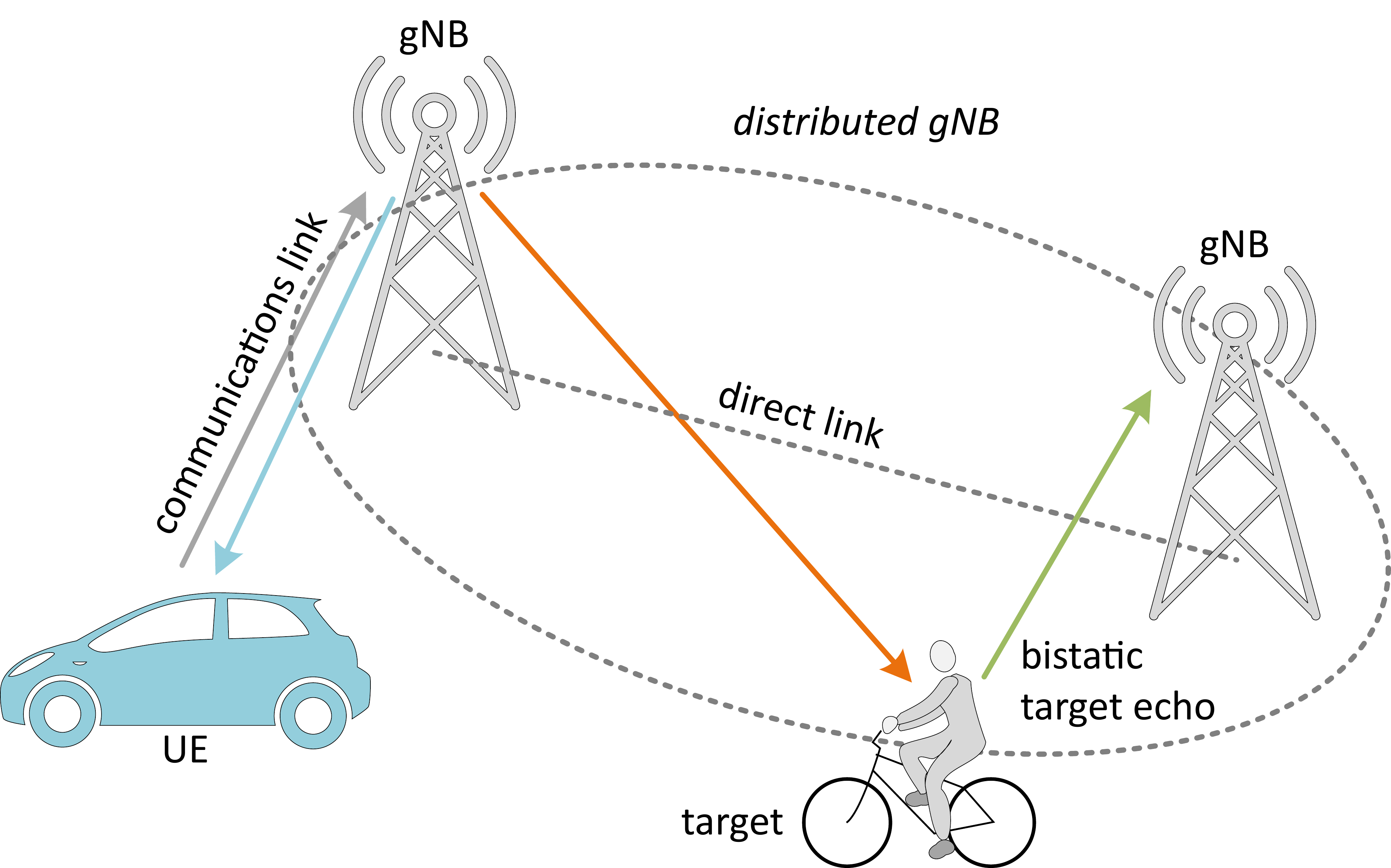}
    \caption{Infrastructure based sensing using distributed radio heads.}
    \label{fig:infrasensing}
\end{figure}

A typical \gls{isac} system consists of either one stand-alone or several meshed radio nodes acting as \gls{tx}, \gls{rx}, or both.
In case of an infrastructure based setup, these can be single or distributed base stations consisting of several synchronized \glspl{rru}.
A single base station case corresponds to a stand alone radar.
The \gls{gnb} must be capable of full duplex radio access and needs to be equipped with an antenna array for \gls{doa} estimation.
The target bearing line will be a circle around the \gls{gnb} and the target location is given by joint \gls{doa} and \gls{tof} (resp. range) estimation, see Fig.~\ref{fig:monostaticing}.
In radar terms, this is referred to as ``monostatic".
The challenge is the full duplex operation of the radio interface, which is not yet standard in communications. The distributed equivalent is depicted in Fig.~\ref{fig:infrasensing}.

\begin{figure}[b]
    \centering
    \includegraphics[width=.83\columnwidth]{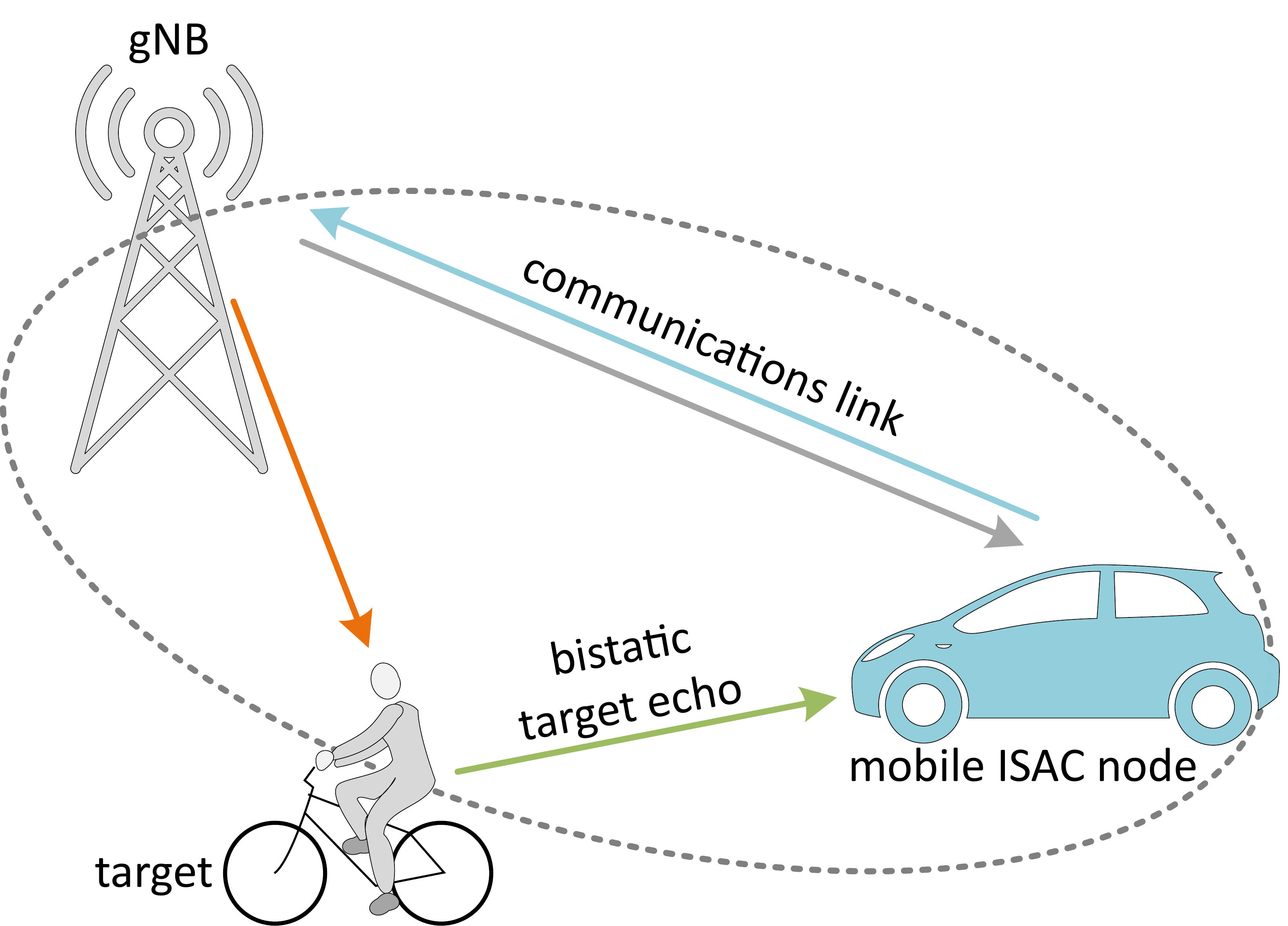}
    \caption{Including the \gls{ue} in the sensing network (\gls{ul}/\gls{dl} sensing).}
    \label{fig:updownsensing}
\end{figure}

It already resembles the passive radar case and, hence, also our \gls{cpcl} scenario \cite{CPCLTho2019}.
The estimated parameter is the excess \Gls{tof} delay of the sensing link relative to the direct \gls{los} link.
The resulting target bearing line is as an ellipse  with the \gls{tx} and \gls{rx} positions as its focal points.
Obviously, we would need multiple measurements, hence additional radio nodes, to achieve a unique and unambiguous location estimate in 2D or even 3D.
\Gls{doa} estimation, hence antenna arrays, are not necessary.
However, beamforming can be additionally applied, e.g., for filtering undesired multipath (clutter).
This distributed \gls{tx}/\gls{rx} geometry is obvious and self-evident for communication centric \gls{isac}.
In radar terms, it is referred to as ``bistatic".
In this case, we do not need a full duplex air interface and we may have further advantages in spatial diversity as will be discussed below.      

The \gls{isac} architecture can also comprise multiple units of mobile \gls{ue} in the \gls{ul} or \gls{dl}, resp., see Fig.~\ref{fig:updownsensing}.
This corresponds to a multiuser scenario, which we call a multisensor scenario in \gls{isac} terminology (the figure shows only one \gls{ue}).
The difference to Fig.~\ref{fig:infrasensing} is that the sensor may now be mobile which has influence on Doppler processing.
Moreover, as the sensor is the \glspl{ue}, the direct wireless link is necessary for \gls{tx}/\gls{rx} synchronization and to make the transmitted signal available at the sensor as a correlation reference.
This is also very close to passive radar, but in contrast to it, the \gls{cpcl} receiver is already prepared by the inherent receiver functionality to generate a clean replica of the transmitted waveform~\cite{CPCLTho2019}.

\begin{figure}[b]
    \centering
    \includegraphics[width=.9\columnwidth]{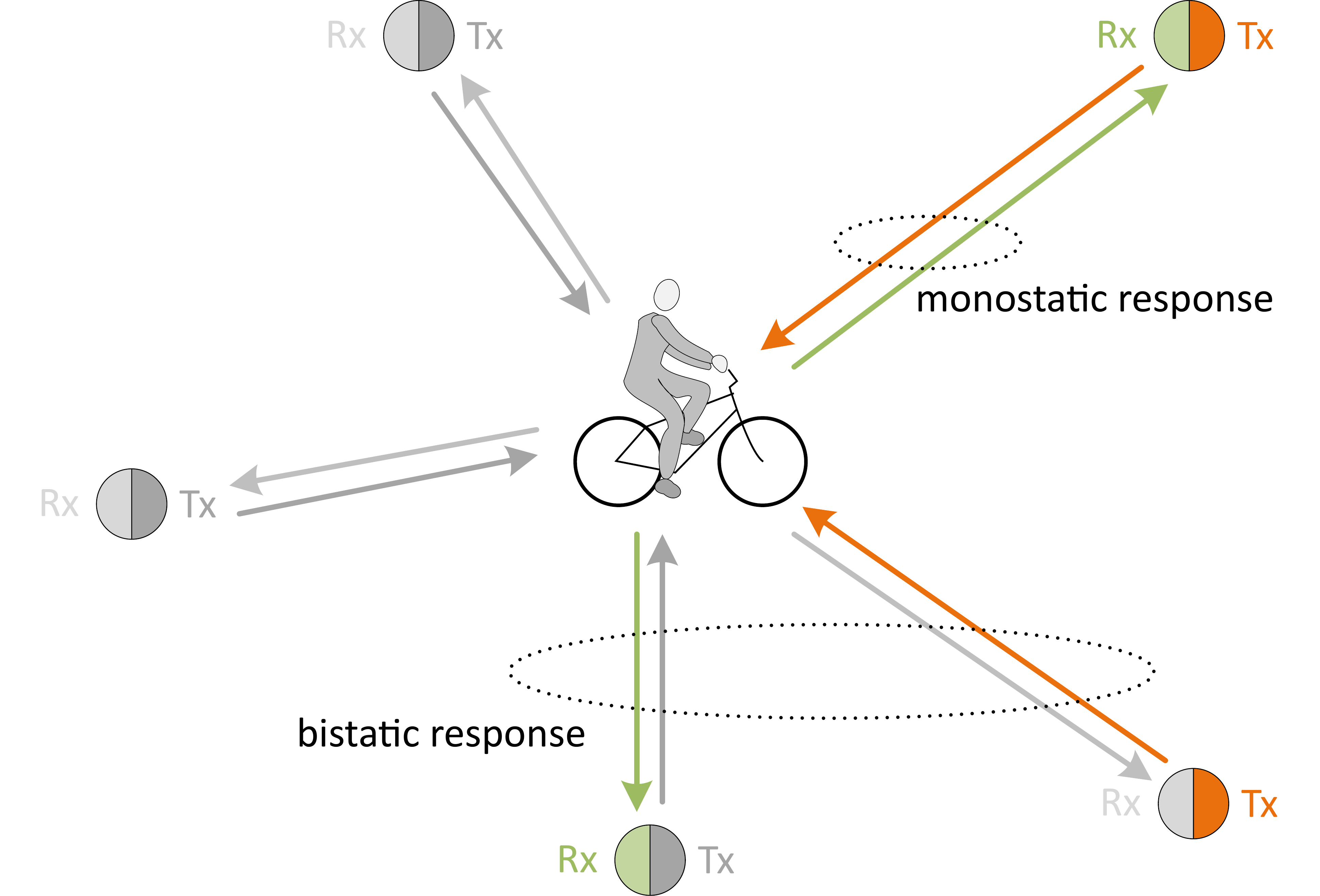}
    \caption{Generic distributed \gls{mimo} radar architecture consisting of multiple transmit and receive links.}
    \label{fig:mimoarch}
\end{figure}

The radar architecture made up from multiple, widely distributed radio nodes is called distributed MIMO radar (as opposed to co-located) \cite{MIMOHai2008}.
The generic distributed \gls{mimo} radar setup shown in Fig.~\ref{fig:mimoarch} involves some issues and challenges.

Obviously, the full $\mathrm{\# Tx} \times  \mathrm{\# Rx}$ \gls{mimo} matrix requires monostatic radio interfaces at all nodes.
Moreover, the multiple \gls{tx}/\gls{rx} links would require some coordinated access, which includes sensor broadcast (with multiple simultaneous \gls{dl} measurements at several \gls{tx}) and the orthogonal multisensor case that can be used for \gls{ul} and \gls{dl} sensing).
Joint transmission can be implemented in the \gls{dl} with noncoherent and coherent superposition at the place of the target.
Moreover, heterogeneous links (including very different frequency bands) can make sense.
However, further discussion is beyond the scope of this paper.
Obviously, distributed \gls{mimo} radar includes several synchronization issues.
It allows unambiguous 3D location and dynamic state vector estimation (which includes the velocity vector and perhaps higher derivatives) if there are enough orthogonal \gls{tx}/\gls{rx} links available.
An excess number of measurements will increase the reliability of detection and estimation.
Reason is that the bistatic view offers a considerable target related backscatter diversity gain.
This applies also to velocity estimation.
While single radar (monostatic or bistatic) is ``Doppler blind" if the target moves tangential to the range bearing line, a distributed geometry can avoid Doppler blindness\cite{TargetHe2010} by combining different measurements.
For dynamic targets, it is important to gather these multiple measurements in a short enough time interval in order not to lose the geometric consistency of the parameters.
Moreover, it seems that Doppler estimation plays a specific role here for target dynamic state estimation, since correct Doppler estimation requires coherent sampling on some slow time interval.
This presumes that the spatial sampling interval should normally be less than half the carrier wavelength.
This is important not only for unambiguous speed estimation.
It is also crucial for achieving a valuable radar integration time in connection with correlation processing.
The resulting \gls{snr} gain is an important advantage of \gls{isac} vs. communication.
While the radar equation seems to penalize \gls{isac} vs. communication with respect to coverage distance, correlation gain can compensate a lot of the additional propagation loss.

\section{Multipath propagation and ISAC Performance Evaluation }
From mobile radio research we already know that the knowledge about propagation phenomena is of highest importance for system design and performance assessment.
For instance, only the idea to exploit rich multipath propagation brought the breakthrough for capacity efficient \gls{mimo} communication.
We have learned that performance evaluation of communication systems on the link and system level has to consider realistic radio propagation models.
Consequentially, advanced sounding and propagation modeling methods were required to provide us with a deeper understanding of the relevant propagation phenomena and to deliver a modeling kernel to be used in more comprehensive system simulation frameworks.
A similar situation we expect for performance assessment of \gls{isac} systems.
While basic \gls{isac} system design approaches assume clear \gls{los} propagation situations, the reality is more complicated.
Propagation phenomena will, for instance, have influence on visibility of targets, as the target can be hidden by obstacles.
Multipath reflections from the environment will further add, which may be considered as some kind of interference (called ”clutter”).
On the other hand, multipath can be exploited to support target illumination and visibility.
Eventually, the scattering statistics of a target is of highest importance for predicting detection probabilities and, hence, for a meaningful information- and estimation-theoretic performance assessment of the whole \gls{isac} system.
Therefore, propagation models for \gls{isac} systems have to reflect already the key \gls{isac} system design issues.
However, we may ask, if the same models that were developed for communications technology can be reused for \gls{isac} or what are the differences and are there specific requirements? 

\begin{figure}[tb]
    \centering
    \includegraphics[width=.8\columnwidth]{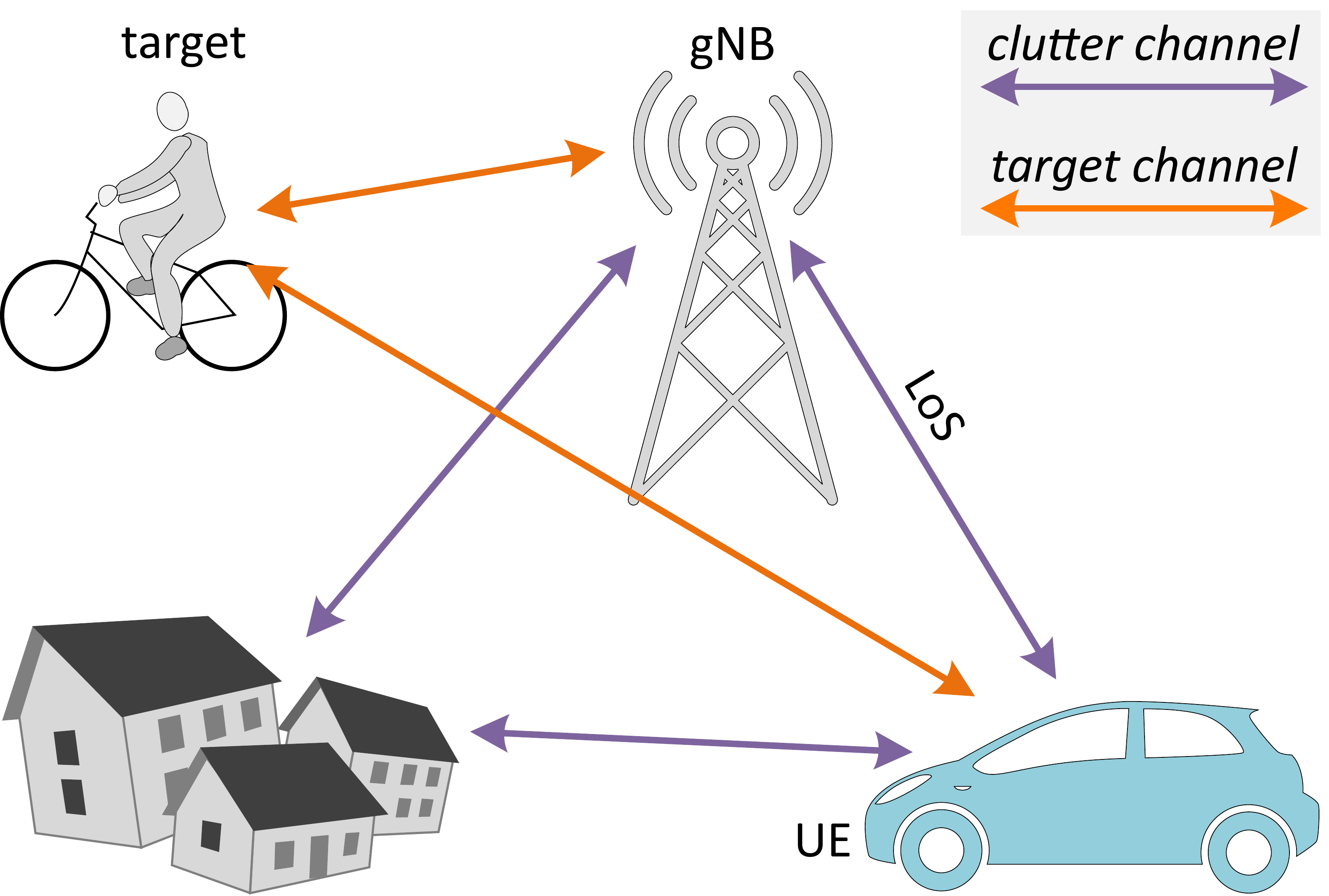}
    \caption{Geometric structure of the \gls{isac} multipath channel consisting on target and clutter paths.}
    \label{fig:geomstruct}
\end{figure}

\begin{figure}[tb]
    \centering
    \includegraphics[width=.9\columnwidth]{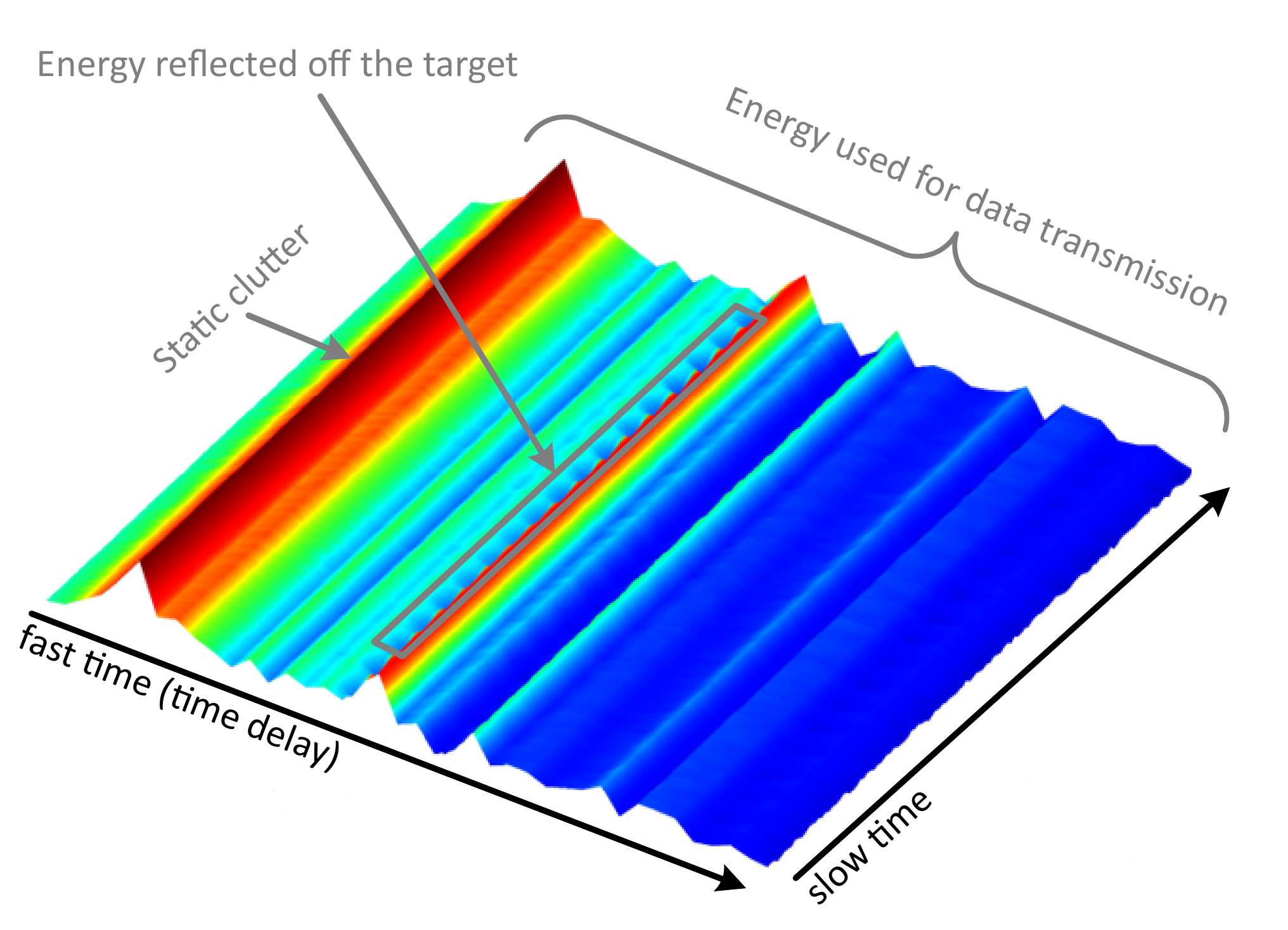}
    \caption{Measured sequence of instantaneous power-delay profiles (1.5 m track, target: car, clutter: buildings and lamp posts).}
    \label{fig:pdp}
\end{figure}

Just like the communication channel, also the sensing channel is composed of multiple propagation paths attributed to direct \gls{los} propagation and deflected paths resulting from interaction with the environment.
However, a communication channel exploits all the energy, which is transmitted from \gls{tx} to \gls{rx} to transport information, regardless on which path it propagates.
In \gls{isac} it is different. Here the \gls{tx}-to-\gls{rx} path that is routed via the target is most interesting as it carries the information about the target's position and dynamics.
The remaining multipath constitutes the clutter, see Fig.~\ref{fig:geomstruct}.
A measured power delay profile (magnitude squared impulse response) is shown as an example in Fig.~\ref{fig:pdp}.
The figure exemplifies how dominating the clutter can be in a practical situation.

\begin{figure}[tb]
    \centering
    \includegraphics[width=.9\columnwidth]{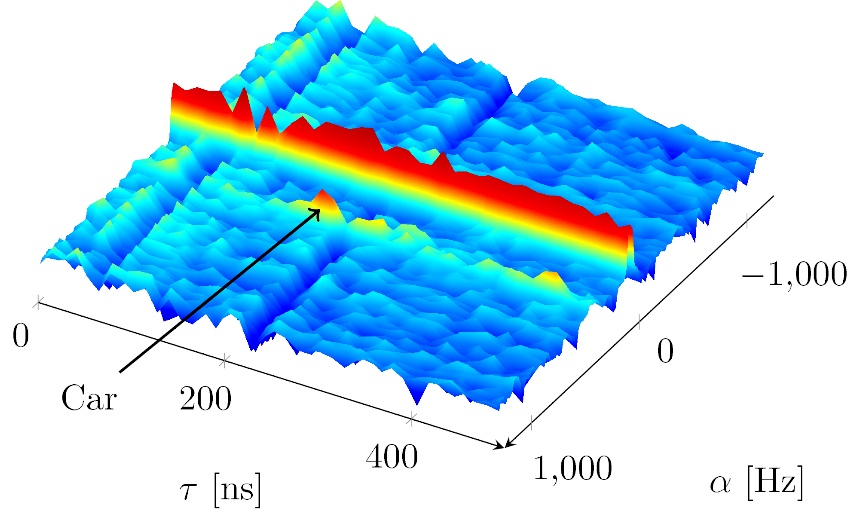}
    \caption{Measured delay-Doppler distribution of a moving car in a static environment calculated from the consecutive complex impulse responses}
    \label{fig:pdp_car}
\end{figure}

The scenario was a street with office buildings at both sides and some lamp posts.
The target was a car sensed via bistatic \gls{los} response from two other cars (\gls{tx} and \gls{rx}, resp.).
The energy reflected off the target almost disappears in the clutter.
As the target car is moving and the clutter is static, the target reveals itself only as some small ripple in Fig.~\ref{fig:pdp}.
Target and clutter can be better separated, if Doppler is resolved by another FFT along the slow time axis.
In Fig.~\ref{fig:pdp_car} the car better reveals as a peak with a discrete Doppler shift corresponding to its speed.
The contribution of the static clutter collapses at zero Doppler. In the same Figure the parameters of two dominant static paths (the direct Tx-to-Rx path and one dominant reflection) have been estimated by a high resolution parameter estimator and subtracted before FFT application.
The separation of clutter vs. target response is a major task of radar signal processing.
The procedures applied typically exploit the different dynamic of the target and the environment by estimating the different Doppler shift or simply apply background subtraction.

Target estimation procedures that proactively exploit the multipath propagation are also interesting.
On the one hand, multipath contributes to target illumination (if there are multiple bounce reflections).
This increases the energy that falls onto the target and brings similar diversity gain as already discussed in relation with Fig.~\ref{fig:mimoarch}.
The potential benefit is a better localization performance.
The target can be localized even if the number of illuminators is too low to deliver the necessary geometrical degrees of freedom.
It becomes also possible to locate objects in shadowed areas \cite{ZetLook2015}, \cite{BlindAlg2007}.
However, the position of the dominant interacting points multilink of the environments need to be known.
An alternative is to apply machine learning from reference scenarios \cite{ImprovSou2021}, \cite{AppSou2019} with known target positions, e.g., for supervised learning.

Yet another approach that exploits multipath relies on target link adaptation.
If we can estimate the \gls{tx}-to-target link, it becomes possible to pre-distort the transmit signal by convolving it with the time axis mirrored channel response.
This method is called time reversal \cite{TimJin2010}.
As a consequence, several copies of the \gls{tx} waveform, which arrive over multiple propagation paths, coincide at the target in time and perhaps also with the same phase (non-coherent and coherent superposition).
This makes sure that all transmitted paths arriving at the target via the multiple paths contribute to the target illumination (effectively focussing and increasing ilumination power) and do not need to be separated as \gls{los} vs. clutter illumination.
An equivalent predistortion approach, related to \gls{otfs} modulation  can be carried out in the Doppler domain, thus matching the Doppler shifts over all illumination paths.
This would considerably decrease the effective target Doppler spread, simplify target estimation and tracking and allow longer coherent radar integration intervals. 

\subsection{Challenges in Multi-link Propagation Modeling}
Since the target state vector (position, orientation and dynamics) is jointly estimated from multiple distributed sensing link parameters, it becomes obvious that the channel model has to be geometrically relevant. 
This is somewhat in contrast to \glspl{gbscm} that are used in mobile radio communications. \Gls{gbscm} represent the multipath geometry only in some statistical sense, e.g., to reproduce joint delay and angular statistics. For \gls{isac} we need deterministic models with more physical relevance. This concerns the correct modelling of incoming and outgoing multipath at the target. If the aim is to exploit multipath, also the deterministic modeling of the dominant reflections is important. Only remaining (not used) clutter can be modeled in a statistical sense. 

Deterministic geometrical relevance includes spatial consistency which means that the geometric parameters that are relevant for target location, change jointly in a consistent way for all multiple Tx/Rx sensor links and over time according to the trajectories of \gls{tx},  \gls{rx}, and target. This is clearly opposed to the random drop based modeling approach applied for a sequence of \gls{gbscm} impulse responses in communications. The consistent, contiguous change of geometry requires a corresponding variation of the parameters of the model which are estimated and tracked by the multiple distributed ICAS sensors. These are the bistatic \gls{tof} and Doppler as well as the directions of propagation paths at both sides of the link (\gls{doa} and \gls{dod}). The consistent change of geometric parameters is encompassed by the dynamic ray tracing approach which was recently proposed in ~\cite{bilibashi_raytracing}. The computational efficient procedure includes the prediction of wave interaction points to minimize the need for repeated full ray tracing search. 

For proper simulation of ISAC signal processing and parameter estimation we need a continuous stream of contiguous channel impulse responses (CIR). While the width of the required parameter grid of \gls{tof}, \gls{dod}, \gls{doa}, and Dopler depends on the available resolution in delay and direction (due to the limited observation aperture in frequency (bandwidth) space (array size)), and slow time (Doppler), the situation with Doppler is somewhat different, though. 
We have to maintain carrier phase coherency over the coherent processing interval along slow time which is used for coherent averaging or Doppler resolution. At the same time the spatial sampling distance should be smaller than half the carrier wavelength. In order to keep up with this we would need an extreme high simulation update rate. Fortunately enough we can generate sequences of impulse responses that keep pace with spatial carrier phase rotation (hence Doppler) from a fixed set of \gls{tof}, \gls{dod}, \gls{doa} path parameters (assuming that these geometric parameters do not change more than the aspired resolution within the coherent processing interval). The equivalent problem we have with sounding measurements. Here we take coherent sequences of impulse responses in slow time intervals that are sampled fast enough to meet the Nyquist criterion in the Doppler phasor domain and long enough to get reliable joint range/Doppler-estimates. 

In total, a propagation channel model for \gls{isac} should be geometrically correct for the multiple simultaneous sensor links and contiguous (differentible) in the parameter space.

\subsection{Challenges in Bistatic Target Reflectivity Modeling}

Besides of the more global, track related view discussed above, we can also have a more local or microscopic view, which is related to the target as a single or solitaire object.
For obvious reasons, the reflectivity of the target object is of outstanding importance for radar performance evaluation\cite{StaMyi2019}.
For \gls{isac} it is the bistatic reflectivity, as discussed with the system setup.
For a comprehensive experimental characterization of a target object we have to consider illumination from all possible directions in azimuth and elevation (practically only in the upper half sphere in case of cares and mostly in lower half sphere in case of flying objects) and observation also from all possible directions.
This way, we end up with a four-dimensional data structure.
Since \gls{tx} (illumination) and \gls{rx} (sensing/observation) angles can take any value, we have the specific cases of \SI{0}{\degree} and \SI{180}{\degree} aspect angles included.
The former corresponds to the monostatic case, whereas the latter is called forward scattering case.
Forward scattering inklcudes shadowing or obstruction of the \gls{los} between Tx and Rx but also diffraction around the target. So it is most distance and frequency dependent. 
However, it is also an interesting radar operation mode as diffraction can increase received power and, hence, enhance target visibility.

In general, the radar target has to be considered as an extended target.
This means that it is wider than a resolution cell, which is given in radial distance by the range resolution.
Therefore, we have to collect multiple range bins for depth information depending on the available bandwidth.
The number of relevant range bins depends on the electrical size of the object, which may be larger than the mechanical one if the target is concave and has structural resonances, e.g., because of cavities.
This way we effectively get another dimension besides of the 4D angles.
Moreover, radial distance of Tx and Rx antennas from the target matters.
In general, we are not in the far field, which is characterized by planar wave fronts along the whole size of the object.
Interestingly enough, this not only true for a practical measurement setup (see next section), but also for many typical \gls{isac} application scenarios in the field, e.g., for car2car communications.
Therefore, a near-to-far-field transform for model building would not suffice.
We would need a model, which is scalable with \gls{tx} and \gls{rx} distance.
Further issues are related to the polarimetric target response, which requires a complete $2 \times 2$ Jones matrix for any entry of the multidimensional radar reflectivity.
We propose the term ``bistatic radar reflectivity” as opposed to bistatic \gls{rcs}, since \gls{rcs} is usually understood as the cross section of an equivalent sphere that reflects the same power as the target in the far field.
The \gls{rcs} is important for radar link budget estimation using the radar equation.
However, the conventional definition of \gls{rcs} does not hold for extended targets.
On the other hand, also the bistatic reflectivity can be calibrated in a way that it reflects the physically correct link budget. 

\begin{figure}[tb]
    \centering
    \includegraphics[width=\columnwidth]{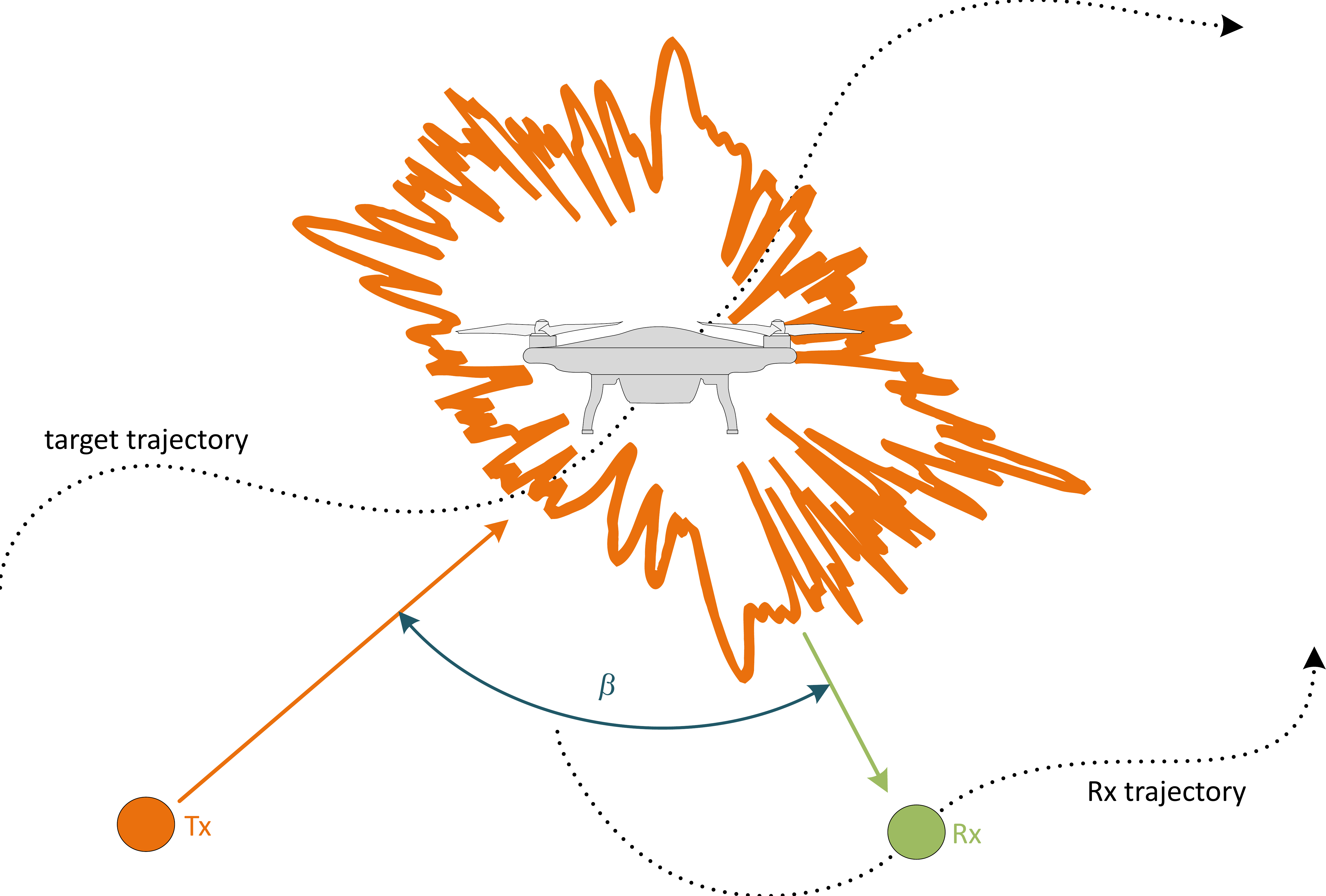}
    \caption{Bi-/Multistatic target reflectivity.}
    \label{fig:reflectivity}
\end{figure}

Fig.~\ref{fig:reflectivity} indicates the bistatic response of an extended target if illuminated and sensed by distributed Tx and Rx antennas.
It becomes obvious that we can attribute a ``global” Doppler shift describing the movement of the target on a track as described in the subsection above.
However, what happens, if the target moves more locally?
If it rotates, or if it contains  rotating parts like wheels or propellers?
Or if the target is a creature like a human being or an animal?
There may be moving legs, arms, or wings.
Or what if the target is tumbling on its track or just starting to change direction?
This time variance would also cause some phase variation over time and, hence, Doppler shift.
But as it is local and independent from the track related to global Doppler, it is better called micro Doppler \cite{MicZha2017},~\cite{chipengo_ai_udoppler}. Related to the discussion about the bistatic reflectivity above, we regard it as the time-variant bistatic reflectivity.
The signal analysis tools to be applied in order to identify and analyze the type and nature of the time variability reach from harmonic analysis over short-time Fourier transform until Wigner distribution and Wigner-Ville spectral analysis, which also includes cyclostationary spectral analysis if the time-variability is not strictly periodic but rather an almost period modulated random process.

\section{Bistatic Target Reflectivity Measurements}

From the discussion above, we conclude that for \gls{isac} propagation research we need two measurement setups.
The first one is a multi-node real-time channel sounder that is capable to emulate a meshed multisensor \gls{isac}-network in a dynamic scenario including multiple mobile radio nodes and moving targets~\cite{MeasBeu2022,ExpSom2018 ,Multinode}.
The multi-link sounder described in~\cite{vtc23sounder} (see also the corresponding TD) can be configured for simultaneous sounding of multiple moving links with cylindrical arrays at Tx and Rx for contiguous recording and joint multi-link DoA/DoD/Doppler estimation.     

\begin{figure}[tb]
    \centering
    \includegraphics[width=.9\columnwidth]{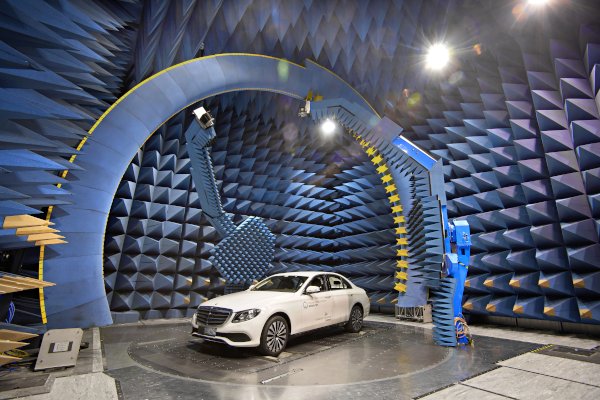}
    \caption{BiRa measurement range for bistatic radar reflectivity at TU Ilmenau (the Stargate antenna arch is not used in this case).}
    \label{fig:BiRa}
\end{figure}

Because of the prominent role of the multistatic target reflectivity in distributed \gls{isac}, we need also a measurement range, being capable to analyze the bistatic radar reflectivity of extended solitaire objects.
First ideas of such a setup were already published in \cite{RoedARC2017}.
Fig.~\ref{fig:BiRa} shows the BiRa measurement range which was recently installed at Ilmenau University of Technology.
It consists of two pivoting gantries that carry a dual polarimetric \gls{tx} and \gls{rx} antenna, resp.
Moreover, a depth sensing camera is installed to record the geometric shape of object under test, which is placed on a turntable.
This, together with the moving gantries, allows independent illumination and observation of the target object from arbitrary directions in the upper half space.
The accessible frequency range covers all relevant frequency ranges from FR1 and FR2 up to \SI{170}{\GHz}.
A remarkable feature is that besides of a standard \gls{vna}, we can use a wide-band 2x2 channel sounder for target illumination and reflectivity sensing with an instantaneous bandwidth of up to \SI{4}{\GHz}.
This does not only accelerate the total measurement time.
It allows also to investigate micro Doppler of time variant targets, e.g., multi-copters with rotating propellers.
Analyzing, studying, and learning the micro Doppler signature of these objects will help to classify their type and to identify and separate them in real-life multiple target scenarios.
The fast and programmable availability of the BiRa measurement range allow to collect training data for AI/ML based target classification and modeling.

\begin{figure}[tb]
    \centering
    \includegraphics[width=.9\columnwidth,trim={6cm 5cm 13.5cm 0.5cm},clip]{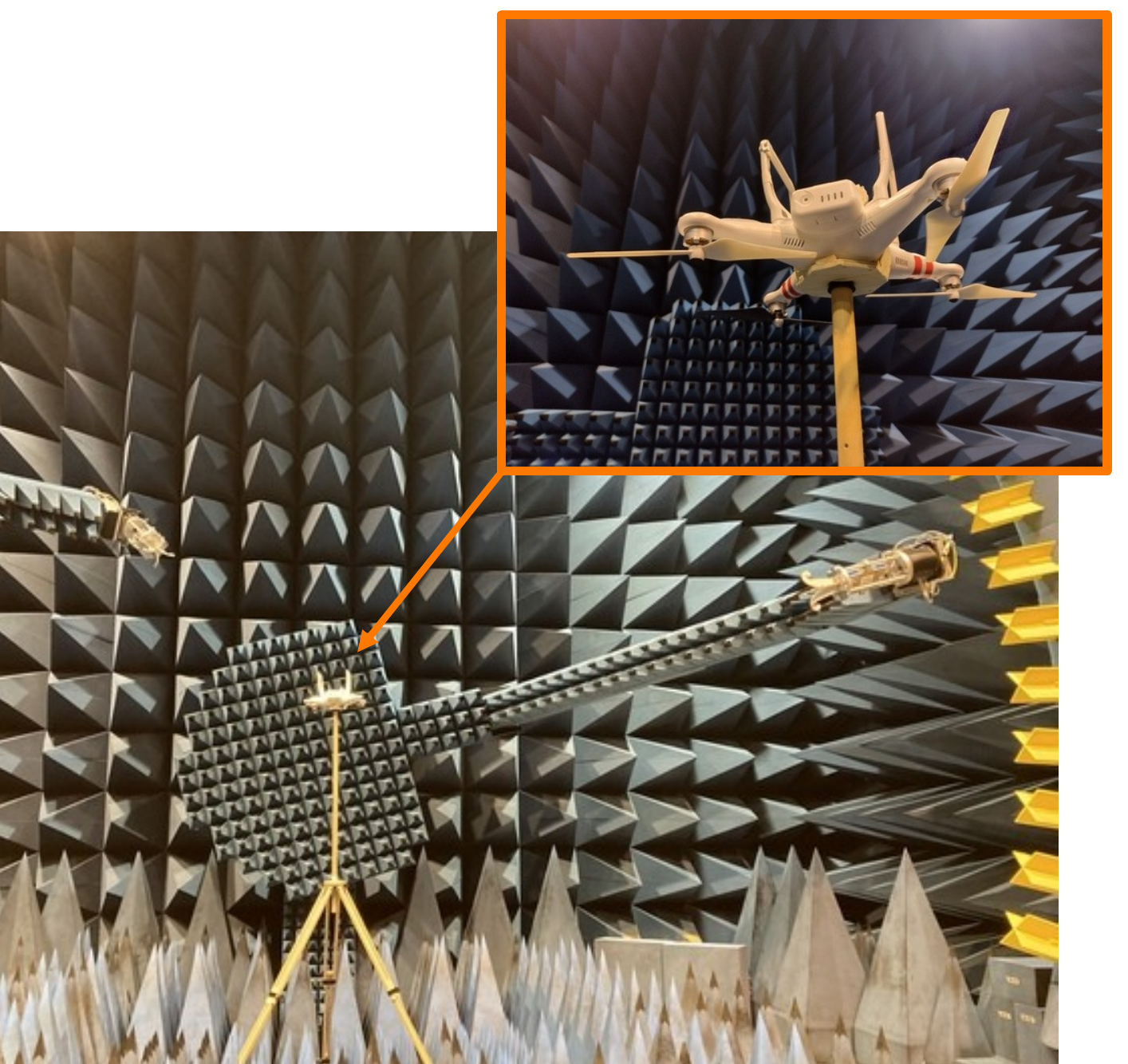}
    \caption{DJI Phantom drone upside down in the BiRa measurement range.}
    \label{fig:dji_phantom}
\end{figure}

In the following paragraph we present two first measured examples that demonstrate some capabilities of BiRa.
Both measurements were related to quad-copters.
The first one was carried out with a DJI Phantom with standing propellers, the second one was made with a bigger quad-copter and rotating propellers.
Fig.~\ref{fig:dji_phantom} depicts the DJI Phantom in the BiRa test range mounted upside down on a wooden tripod. 
The two gantries were moved over the drone in such a way that a drone flyover over two bistatic radio nodes (Tx and Rx) placed in a row along the ground track was emulated.

\begin{figure}[tb]
    \centering
    \includegraphics[width=.9\columnwidth]{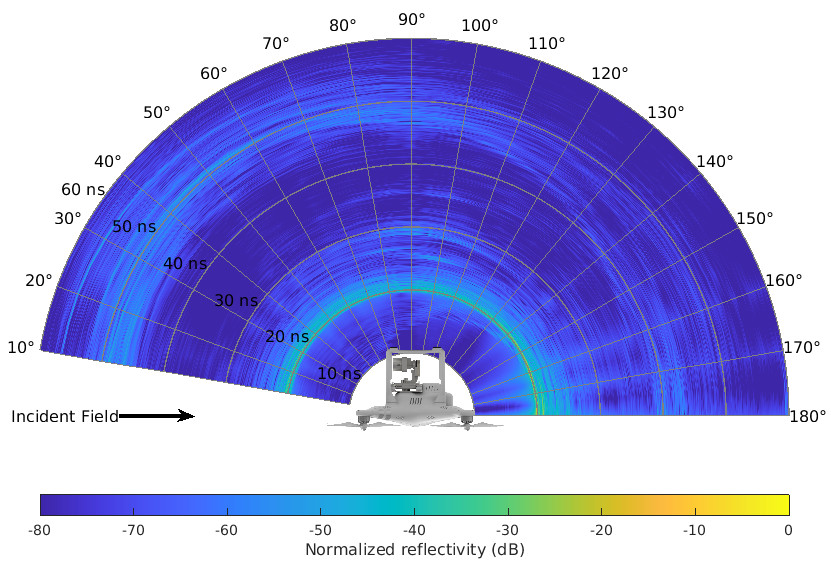}
    \caption{Bistatic Radar reflectivity of DJI Phantom measured over 170 deg bistatic azimuth angle span (background removal applied).} 
    \label{fig:extended_target}
\end{figure}

\begin{figure}[tb]
    \centering
    \includegraphics[width=.9\columnwidth]{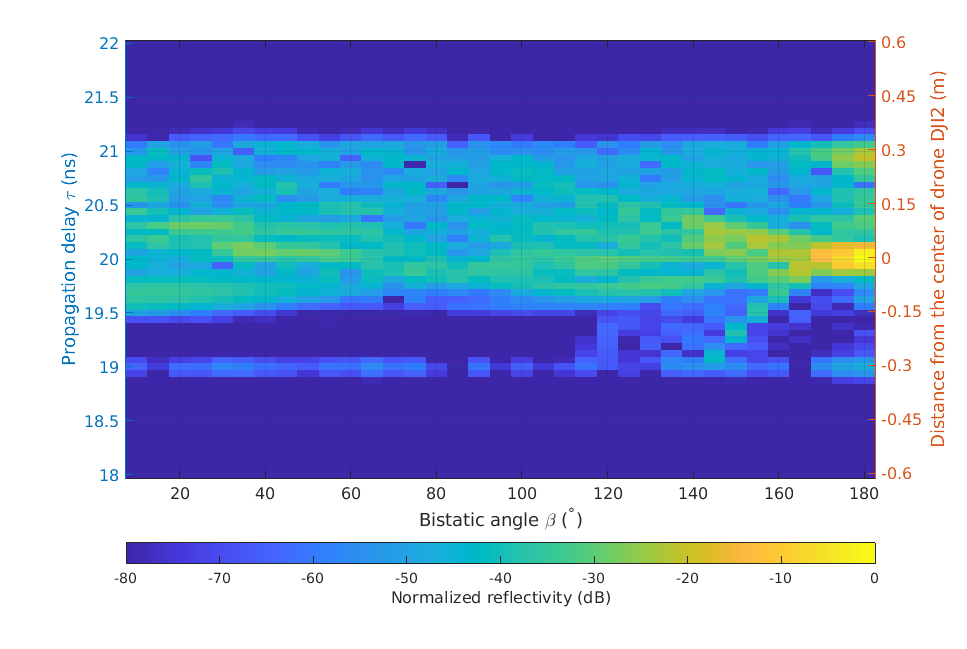}
    \caption{Bistatic Radar reflectivity of DJI Phantom measured within a target window of 2ns over 170 deg bistatic azimuth angle span. } 
    \label{fig:extended_target zoom}
\end{figure}

One gantry was fixed relative to the drone (on same elevation level as the drone), the other gantry was moving from 10 to 180 deg relative to the fixed gantry, thus emulating a flyover path starting from far away (where the two radio nodes appear as quasi-monostatic) until a situation where the drone is just in between the two nodes, which is called the forward scattering case.
The Tx and Rx polarization was horizontal. 
The measured data were recorded with an VNA over a frequency span from 2 to 18 GHz.
Parasitic reflections and the direct LoS Tx-to-Rx cross talk were removed by background subtraction and time gating.
The latter is supported by the wide bandwidth of 16 GHz.
For characterization of the target reflectivity in application specific bands the bandwidth can be further reduced and matched to the respective target system parameters.
Besides of the half circle track that corresponds to the target, remaining parasitic reflections (after background subtraction) can be seen in Fig.~\ref{fig:extended_target}.
Fig.~\ref{fig:extended_target zoom} depicts contributions that are attributed to the target within a time window of 2 ns width, which corresponds to the 30x30 cm size of the drone. 
For the first part of the track we observe a delay spread of the target of abundant 1.5 ns.
At the end of the track we are more approaching the forward scattering case.
Therefore, the delay spread should become smaller, as can be seen in the figure.
However, the more we approach the 180 deg case, a parasitic component with delay of 1 ns shows up corresponding to the mounting plane of the antenna which is in a distance of ca. 15 cm behind the phase center of the antenna. 
The parasitic effects need to be investigated in more detail and mitigation methods are necessary.
Obviously, the forward scattering case with Tx and Rx antennas in opposition to each other is the most difficult case.                               

\begin{figure}[tb]
    \centering
    \includegraphics[width=.9\columnwidth,trim={0px 0px 0px 38px},clip]{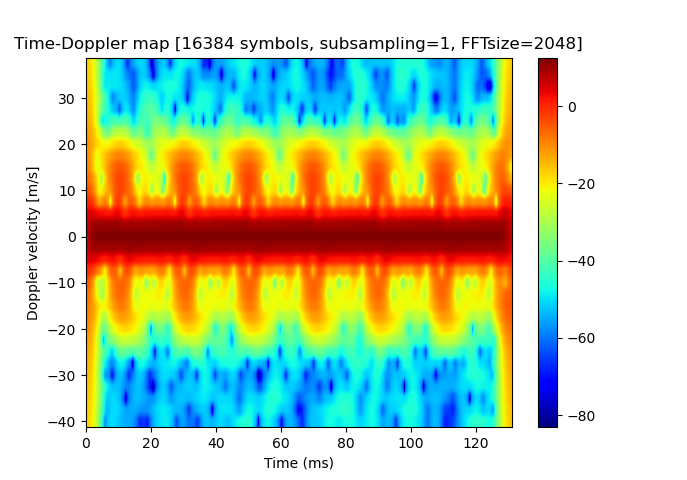}
    \caption{Micro-Doppler signature of single drone propeller}
    \label{fig:micro_doppler}
\end{figure}

The second example shows the backscattering of a time variant target, hence micro-Doppler.
A quadcopter was again used as a test object.
Here, only one propeller was rotating.
It was illuminated by the Tx in the horizontal plane (0 deg elevation) and sensed by the Rx from 30 deg elevation (here from above since this corresponds to -30 deg from below if the drone were flying) and 90 deg separated in azimuth.
The used bandwidth was \SI{160}{\MHz} (1280 carriers), centered at \SI{3.7}{\GHz}, OFDM symbol length \SI{8}{\micro\second}.
The Doppler spectrum was calculated by a short time Fourier transform sliding over slow time from a consecutive sequence of OFDM symbols (FFT 2048 samples, 32 samples window sliding step, Gaussian window).
The resulting spectrogram is shown in Fig.~\ref{fig:micro_doppler}.
The slow time axis indicates a 20 ms period.
It is interesting to see that the Doppler spectrum is continuous, despite of the constant rotation rate (rpm).
This is not a surprise, since the the propeller velocity increases linearly toward the tips.
Therefore, other, more sophisticated signal processing schemes may be necessary to identify the the rotation rate from the Doppler spectrum.

\section{Conclusions}
Starting with a comprehensive overview on the architectural concepts of distributed multi-sensor ISAC (see also \cite{icas_multisensor}) we have worked out the requirements for dynamic channel modeling and measurement. It turns out that we need a simulation framework for multipath propagation, which is capable to generate spatially consistent and geometrically representative multiple link responses. Thereby it seems that phase-continuous modeling and correct bistatic reflectivity modeling of moving targets poses new challenges.

The same applies to measurement technology. Here, we need dynamic multi-node sounding devices with directional resolution capabilities that are able to follow the Doppler phasor rotation over sufficiently long tracks in real-time. In addition, because of the outstanding importance of the target response, we need another measurement range that allows us to study the bistatic target reflectivity of spatially extended time-varying solitary targets. The results gained with the BiRa measurement range will allow to deduce sophisticated scattering models of the target to be "plugged into" in a more comprehensive dynamic propagation simulation framework (dynamic ray tracing or dynamic GBSCM). The models shall be distance independent (applicable for any distance between Tx/Rx and target) and easy to be traced by  higher-level multipath simulation framework. Because of the deterministic nature of localization and target recognition, these models will need a deterministic component which is related to the shape of target. However, we may also need a stochastic part, in order to balance the levels of detail, effort, and generalisation. Therefore, the long lasting discussion about stochastic vs. deterministic modeling of propagation phenomena will get a second wind -- related to target modeling in multisensor ISAC.

At the same time, the excessive and fast measurement capacity of the BiRa testbed enables the generation of a huge amount of data for statistical target classification and model building, which can be used as a digital target twin for machine learning and AI/ML-based ISAC scenario recognition.

\section*{Acknowledgments}
The research has been funded by the Federal State of Thuringia, Germany, and the European Social Fund (ESF) under grants 2016 FGR 0039 (project ``KoSiMoLo"), 2017 FGI 0007 (project ``BiRa"), and 2021 FGI 0007 (project ``Kreatör") as well as by the Federal Ministry of Education and Research of Germany under grant 16KISK015 (project ``Open6GHub").

\IEEEtriggeratref{10}
\bibliographystyle{IEEEtran}
\bibliography{References}

\begin{thebibliography}{10}
\providecommand{\url}[1]{#1}
\csname url@samestyle\endcsname
\providecommand{\newblock}{\relax}
\providecommand{\bibinfo}[2]{#2}
\providecommand{\BIBentrySTDinterwordspacing}{\spaceskip=0pt\relax}
\providecommand{\BIBentryALTinterwordstretchfactor}{4}
\providecommand{\BIBentryALTinterwordspacing}{\spaceskip=\fontdimen2\font plus
\BIBentryALTinterwordstretchfactor\fontdimen3\font minus
  \fontdimen4\font\relax}
\providecommand{\BIBforeignlanguage}[2]{{%
\expandafter\ifx\csname l@#1\endcsname\relax
\typeout{** WARNING: IEEEtran.bst: No hyphenation pattern has been}%
\typeout{** loaded for the language `#1'. Using the pattern for}%
\typeout{** the default language instead.}%
\else
\language=\csname l@#1\endcsname
\fi
#2}}
\providecommand{\BIBdecl}{\relax}
\BIBdecl

\bibitem{CPCLTho2019}
R.~S. Thoma, C.~Andrich, G.~D. Galdo, M.~Dobereiner, M.~A. Hein, M.~Kaske,
  G.~Schafer, S.~Schieler, C.~Schneider, A.~Schwind, and P.~Wendland,
  ``{Cooperative Passive Coherent Location: A Promising 5G Service to Support
  Road Safety},'' \emph{IEEE Communications Magazine}, vol.~57, no.~9, pp.
  86--92, 2019.

\bibitem{JCASOverviewTho2021}
R.~Thomä, T.~Dallmann, S.~Jovanoska, P.~Knott, and A.~Schmeink, ``{Joint
  Communication and Radar Sensing: An Overview},'' in \emph{2021 15th European
  Conference on Antennas and Propagation (EuCAP)}, 2021, pp. 1--5.

\bibitem{MicZha2017}
Q.~Zhang, Y.~Luo, and Y.~Chen, \emph{{Micro-Doppler Characteristics of Radar
  Targets}}.\hskip 1em plus 0.5em minus 0.4em\relax Elsevier, 11 2016.

\bibitem{MIMOHai2008}
A.~M. Haimovich, R.~S. Blum, and L.~J. Cimini, ``{MIMO Radar with Widely
  Separated Antennas},'' \emph{IEEE Signal Processing Magazine}, vol.~25,
  no.~1, pp. 116--129, 2008.

\bibitem{TargetHe2010}
Q.~He, R.~S. Blum, H.~Godrich, and A.~M. Haimovich, ``{Target Velocity
  Estimation and Antenna Placement for MIMO Radar With Widely Separated
  Antennas},'' \emph{IEEE Journal of Selected Topics in Signal Processing},
  vol.~4, no.~1, pp. 79--100, 2010.

\bibitem{ZetLook2015}
R.~Zetik, M.~Eschrich, S.~Jovanoska, and R.~S. Thoma, ``{Looking behind a
  corner using multipath-exploiting UWB radar},'' \emph{IEEE Transactions on
  Aerospace and Electronic Systems}, vol.~51, no.~3, pp. 1916--1926, 2015.

\bibitem{BlindAlg2007}
\BIBentryALTinterwordspacing
V.~Algeier, B.~Demissie, W.~Koch, and R.~Thomä, ``{State Space Initiation for
  Blind Mobile Terminal Position Tracking},'' \emph{EURASIP Journal on Advances
  in Signal Processing}, vol. 2008, no.~1, p. 394219, Oct 2007. [Online].
  Available: \url{https://doi.org/10.1155/2008/394219}
\BIBentrySTDinterwordspacing

\bibitem{ImprovSou2021}
\BIBentryALTinterwordspacing
M.~N. de~Sousa, R.~Sant´Ana, R.~P. Fernandes, J.~C. Duarte, J.~A. Apolinário,
  and R.~S. Thomä, ``{Improving the performance of a radio-frequency
  localization system in adverse outdoor applications},'' \emph{EURASIP Journal
  on Wireless Communications and Networking}, vol. 2021, no.~1, p. 123, May
  2021. [Online]. Available: \url{https://doi.org/10.1186/s13638-021-02001-6}
\BIBentrySTDinterwordspacing

\bibitem{AppSou2019}
M.~N. de~Sousa and R.~S. Thomä, ``{Applying Random Forest and Multipath
  Fingerprints to Enhance TDOA Localization Systems},'' \emph{IEEE Antennas and
  Wireless Propagation Letters}, vol.~18, no.~11, pp. 2316--2320, 2019.

\bibitem{TimJin2010}
Y.~Jin, J.~M.~F. Moura, and N.~O'Donoughue, ``{Time Reversal in Multiple-Input
  Multiple-Output Radar},'' \emph{IEEE Journal of Selected Topics in Signal
  Processing}, vol.~4, no.~1, pp. 210--225, 2010.

\bibitem{bilibashi_raytracing}
D.~Bilibashi, E.~M. Vitucci, and V.~Degli-Esposti, ``{On Dynamic Ray Tracing
  and Anticipative Channel Prediction for Dynamic Environments},'' \emph{IEEE
  Transactions on Antennas and Propagation}, pp. 1--1, 2023.

\bibitem{StaMyi2019}
S.~J. Myint, C.~Schneider, M.~Röding, G.~D. Galdo, and R.~S. Thomä,
  ``{Statistical Analysis and Modeling of Vehicular Radar Cross Section},'' in
  \emph{2019 13th European Conference on Antennas and Propagation (EuCAP)},
  2019, pp. 1--5.

\bibitem{chipengo_ai_udoppler}
U.~Chipengo, A.~P. Sligar, S.~M. Canta, M.~Goldgruber, H.~Leibovich, and
  S.~Carpenter, ``{High Fidelity Physics Simulation-Based Convolutional Neural
  Network for Automotive Radar Target Classification Using Micro-Doppler},''
  \emph{IEEE Access}, vol.~9, pp. 82\,597--82\,617, 2021.

\bibitem{MeasBeu2022}
\BIBentryALTinterwordspacing
J.~Beuster, C.~Andrich, M.~Döbereiner, S.~Schieler, M.~Engelhardt,
  C.~Schneider, and R.~Thomä, ``{Measurement Testbed for Radar and Emitter
  Localization of UAV at 3.75 GHz},'' 2022. [Online]. Available:
  \url{https://arxiv.org/abs/2210.07168}
\BIBentrySTDinterwordspacing

\bibitem{ExpSom2018}
G.~Sommerkorn, M.~Käske, D.~Czaniera, C.~Schneider, G.~Del~Galdo, R.~S.
  Thomä, and M.~Walter, ``{Experimental and Analytical Characterization of
  Time-Variant V2V Channels in a Highway Scenario},'' in \emph{2019 13th
  European Conference on Antennas and Propagation (EuCAP)}, 2019, pp. 1--5.

\bibitem{Multinode}
S.~Zelenbaba, B.~Rainer, M.~Hofer, D.~Löschenbrand, A.~Dakić, L.~Bernadó,
  and T.~Zemen, ``{Multi-Node Vehicular Wireless Channels: Measurements, Large
  Vehicle Modeling, and Hardware-in-the-Loop Evaluation},'' \emph{IEEE Access},
  vol.~9, pp. 112\,439--112\,453, 2021.

\bibitem{vtc23sounder}
D.~Stanko, M.~Döbereiner, G.~Sommerkorn, D.~Czaniera, C.~Andrich,
  C.~Schneider, S.~Semper, A.~Ihlow, and M.~Landmann, ``{Time Variant
  Directional Multi-Link Channel Sounding and Estimation for V2X},'' in
  \emph{VTC2023-Spring}, 6 2023.

\bibitem{RoedARC2017}
M.~Röding, G.~Sommerkorn, S.~Häfner, A.~Ihlow, S.~Jovanoska, and R.~S.
  Thomä, ``{ARC\textsuperscript{2}4 — A double-arch positioner for bistatic
  RCS measurements with four degrees of freedom},'' in \emph{2017 47th European
  Microwave Conference (EuMC)}, 2017, pp. 1273--1276.

\bibitem{icas_multisensor}
R.~Thomä and T.~Dallmann, ``{Distributed ISAC Systems – Multisensor Radio
  Access and Coordination},'' in \emph{European Microwave Week (2023), Focused
  Session "Joint Communication and Radar Sensing - a step towards 6G"}, 9 2023.

\end{thebibliography}

\end{document}